\documentclass[twocolumn,superscriptaddress,showpacs,amsmath,amssymb]{revtex4}

\usepackage{graphicx}
\usepackage{amsmath}
\usepackage{bm}

\begin{document}

\title{Surface-mediated attraction between colloids}
\author{H. Ka\"{\i}di}     
\affiliation{Laboratoire de Physique des Polym\`{e}res et Ph\'{e}nom\`{e}nes Critiques, \\
Facult\'{e} des Sciences Ben M'sik, B.P. 7955, Casablanca, Morocco }
\author{T. Bickel}      
\altaffiliation[Electronic adress: ]{th.bickel@cpmoh.u-bordeaux1.fr}
\affiliation{Centre de Physique Mol\'{e}culaire Optique et Hertzienne, Universit\'{e} Bordeaux 1, \\
351 cours de la Libération, 33405 Talence, France}
\author{M. Benhamou}
\affiliation{Laboratoire de Physique des Polym\`{e}res et Ph\'{e}nom\`{e}nes Critiques, \\
Facult\'{e} des Sciences Ben M'sik, B.P. 7955, Casablanca, Morocco }


\date{\today}

\begin{abstract}
We investigate the equilibrium properties of a colloidal solution in contact with
a soft interface. As a result of symmetry breaking, 
surface effects are generally prevailing in confined colloidal systems.
In this Letter, particular emphasis is given to surface fluctuations
and their consequences on the  local (re)organization of the suspension.
It is shown that particles experience a significant effective interaction in the vicinity of the interface.  
This potential of mean force is always attractive, with range controlled by the 
surface correlation length.
We suggest that, under some circumstances, surface-induced attraction may have a strong influence on 
the local particle distribution.
\end{abstract}

\maketitle

Colloidal suspensions are solutions of fairly large objects, with typical size ranging from $1$ nm to $1$ 
$\mu$m. The primary question of their stability 
and phase behaviour is the foundation of many technological applications~\cite{russelbook}. 
Formally, the statistical description of a
colloidal dispersion involves colloid-colloid, colloid-solvent and solvent-solvent
interactions. However, such detailed and complex information is usually not required to understand 
essential features, and it has been found more appropriate to develop effective descriptions 
where the colloids interact through a \emph{potential of mean force}~\cite{likosPR01}.
The individual forces acting between particles then depend
explicitly on the temperature and on the chemical potential of the solvent. Examples of such effective
potentials include dispersion forces, DLVO theory for charged systems, or 
depletion interactions in polydispersed solutions. 

On the other hand,
it has been recognized long ago that surface effects
are prevailing in confined colloidal systems~\cite{aoJCP54,joanny79}. The mutual influence 
of bulk and surface properties on each other is a challenging problem, 
that conversely may lead to unusual behaviours.
For instance, when a bidispersed hard-sphere suspension is brought in contact with a flat substrate,
excluded-volume effects are known to push the larger beads toward the wall
of the sample~\cite{kaplanPRL94}. Recent experiments done with curved or
corrugated surfaces have shown that \emph{geometric} features of the surface can
also create and modulate entropic force fields~\cite
{dinsmoreNature96}. These depletion forces can be used to grow
oriented colloidal crystal, with numerous potential applications such as the
fabrication of photonic bandgap crystals~\cite{linPRL00}. 

In this Letter, we present some new findings regarding the static 
organization of nanoparticles near a \emph{fluctuating} surface.
Adsorption of colloidal particles on a flexible interface is
an essential step in many biological processes, and the underlying physics of this
mechanism has been extensively studied for simple model systems~\cite{lipowskyEPL98,divetEPL02,
desernoEPL03,vaneffenEPL04}.
Thermal undulations of the surface are nevertheless disregarded in most theoretical works, 
even though they are expected to locally alter the particle distribution~\cite{bickelJCP03,bickelPRE04}.
Technically, the difficulty lies in the interplay between
bulk and surface degrees of freedom.
This question is considered hereafter, with particular 
emphasis given to indirect interactions
between colloids that merge from surface fluctuations.
Our aim is to bring down the complexity of the problem and, ultimately, to come down to
an effective description for the colloidal suspension.

The remaining of the presentation proceeds as follows. First, we derive general expressions
for the one and two--body potentials. We then express the local density profile and the corresponding 
adsorbance. The computation of the effective pair-potential for relevant 
experimental configurations is presented in the next section. 
Finally, we draw some concluding remarks 
on whether surface fluctuations could induce colloid crystallization.

The physical system under consideration is a dilute colloidal suspension in
contact with a fluid interface, defined here as a sharp 
boundary. The results derived in this article are intended
to be as general as to describe a wide class of soft surfaces, ranging from
liquid-liquid interfaces to surfactant monolayers or bilayers. 
We assume that the surface deviates only slightly
from the horizontal plane. In the Monge representation, a point of the 
corrugated surface
has coordinates $\left( x,y,h\left( x,y\right) \right) $. The height
function $h\left( x,y\right) $ may take either positive or negative values.
In what follows, we shall use the notation ${\bm r}=\left( {\bm\rho }
,z\right) $, where ${\bm\rho }=\left( x,y\right) $ is the transverse vector
and $z$ the perpendicular distance.
The total Hamiltonian $\mathcal{H}$ of the system can be written as the sum of three
terms 
\begin{equation}
\mathcal{H} =\mathcal{H}_{m} + \mathcal{H}_{cc} + \mathcal{H}_{cm} \ .
\label{hamiltonian}
\end{equation}
The first contribution is the energy of the weakly curved surface~\cite{safranbook} 
\begin{equation}
\mathcal{H}_{m}\left[ h\right] =\frac{1}{2}\int d^2{\bm\rho }\left[ \kappa
\left( \Delta h\right) ^{2}+\sigma \left( \nabla h\right) ^{2}+\mu h^{2}
\right] \ .
\label{helfrich}
\end{equation}
The parameter $\mu$, the  bending rigidity $\kappa$ and the surface tension $\sigma$ are the bare elastic
constants of the interface, in the absence of particle. 
It is convinient to define the expectation value 
$\left\langle X\right\rangle _{0}$ of a given functional $X\left[ h\right] $ as 
\begin{equation}
\left\langle X\right\rangle _{0}=\frac{1}{{\mathcal Z}_{0}} \int \mathcal{D}hX\left[ h\right]
e^{-\beta \mathcal{H}_{m}\left[ h\right] } \ ,
\label{defmean}
\end{equation}
with ${\mathcal Z}_{0}=\int \mathcal{D}h \exp \left[-\beta \mathcal{H}_{m}\left[ h\right] \right]$
the partition function of the interface. We adopt
the usual notation $\beta =1/k_{B}T$,
with $T$ the absolute temperature and $k_{B}$ the Boltzmann constant.
We also define the height correlation function -- or Green function -- of the surface
\begin{equation}
G\left( {\bm\rho }-{\bm\rho }'\right)  =\left\langle h\left( {\bm
\rho }\right) h\left( {\bm\rho }'\right) \right\rangle
_{0}-\left\langle h\left( {\bm\rho }\right) \right\rangle _{0}\left\langle
h\left( {\bm\rho }'\right) \right\rangle _{0}  \ ,
\label{defprop}
\end{equation}
from which we extract the mean-squared displacement $\xi_{\perp}^2=G(0)$.

The direct colloid--colloid interaction $\mathcal{H}_{cc}$ does not need to be specified at this point.
The last term in eq.~(\ref{hamiltonian}), $\mathcal{H}_{cm} $, represents the
colloid-interface interaction. It is generally a
complicated function of particle position and surface configurations.
However, since we are interested in the regime where the interface undergoes strong fluctuations, 
we restrict the discussion to colloids with size much smaller than $\xi_{\perp}$. 
Typically, the particles under consideration have diameter of a few tens of nanometers, 
whereas the surface roughness lies in the micrometer range.
This assumption allows us to neglect finite-size effects and
to select a \emph{local} potential, that depends only on the relative perpendicular 
distance between the colloid and the surface. Despite this simplification, $\mathcal{H}_{cm} $ 
is still expected to be quite complicated. We next assume that the surface potential is
``short-ranged'' (in comparison to the length scale $\xi_{\perp}$), with 
typical extension of the order of the colloidal size. 
We then choose the following contact potential
\begin{equation}
\beta \mathcal{H}_{cm}\left[ h\right] =-\frac{ \omega }{2}\sum_{i=1}^{N}\delta \left( z_{i}-h\left( 
{\bm\rho }_{i}\right) \right) \ ,
\label{intcm}
\end{equation}
with $\delta$ the Dirac distribution. In this definition, the discrete sum runs 
over all particles with position ${\bm r}_i = ({\bm \rho}_i,z_i)$, and 
$\omega > 0$ is the (surface) coupling constant. Actually, $\omega$ plays the role
of an \emph{extrapolation length} as  usually
encountered in surface critical phenomena~\cite{binderbook}.
In this model, the surface is penetrable and the colloids can accomodate on
both sides of the interface. 
It is clear that the potential defined in eq.~({\ref{intcm})
is rather different from usual DLVO or hydratation potentials.
Nevertheless, we focus in the following on colloids that are \emph{slightly bound} to the surface,
so that we do not expect the microscopic details of the potential to be pertinent.
If necesseray, the strength of the attraction $\omega$ may be related to the depth $U_0$
and the range $b$ of a more realistic potential through the relation $k_BT \omega = U_0b$.

To derive the statistical quantities of interest, we first evaluate the 
grand canonical partition function 
\begin{equation}
{\mathcal Z}_{G} = \sum_{N=0}^{+\infty} \frac{f^N}{\lambda ^{3N}N!}\int \prod_{i=1}^N d{\bm r}_{i}
\int \mathcal{D}he^{-\beta \mathcal{H}\left[ h\right] }  
\ ,
\label{gcanonical}
\end{equation}
whith $\lambda $ the thermal wavelength of the particles under consideration,
and $f$ the fugacity. The functional 
integral extends over all conformations of the field $h(x,y)$. The contribution $N=0$ 
is the partition function $\mathcal{Z}_0$ of an interface in a particle-free environment. 
The term $N=1$ corresponds to an interface interacting with one colloid, and so on.
What makes the evaluation of eq.~(\ref{gcanonical}) difficult is 
the fact that bulk and surface degrees of freedom are coupled
through $\mathcal{H}_{cm}$. 
However, for the Gaussian theory considered here, 
surface undulations can be traced out
by using the standard cumulant method~\cite{itzyksonbook}.  
After some algebra, we find that the colloids interact through an effective potential
$\mathcal{H}_{cc}^{\mathit{eff}}=\mathcal{H}_{cc}+ \mathcal{U}$. 
The potential of mean force $\mathcal{U}$ involves many--body interactions
\begin{equation}
\begin{split}
\mathcal{U}\left( {\bm r}_{1},...,{\bm r}_{N}\right) &=-k_{B}T\ln
\left\langle e^{-\beta \mathcal{H}_{cm}\left[ h\right] }\right\rangle _{0}  \\
& = \sum_{i=1}^{N}
\mathcal{U}_{1}\left( {\bm r}_{i}\right) +\sum_{\left\{ i,j\right\} }
\mathcal{U}_{2}\left( {\bm r}_{i},{\bm r}_{j}\right) +... \ .
\label{defmeanforce}
\end{split}
\end{equation}
As we are only interested in the \emph{weak adsorption regime},
contributions from three--body terms and beyond can be neglected if $\omega$ is
small enough. The one and two--body potentials are respectively
\begin{equation}
\begin{split}
&\beta \mathcal{U}_{1}\left( {\bm r}\right) =-\frac{\omega }{2}\Phi _{1}\left(
z\right)\ , \ \mbox{and}  \\
& \beta \mathcal{U}_{2}\left( {\bm r},{\bm r}'\right) =-\frac{\omega ^{2}}{8
}\left[ \Phi _{2}\left( {\bm\rho },{\bm\rho }';z,z'\right)
-\Phi _{1}\left( z\right) \Phi _{1}\left( z^{\prime }\right) \right] \ ,
\label{u2}
\end{split}
\end{equation}
where we have introduced the useful function 
$\Phi _{N}( {\bm r }_{1},...,{\bm r }_{N})
  =  \big\langle \prod_{i=1}^{N}\delta ( z_{i}-h( {\bm\rho }
_{i}) ) \big\rangle _{0} $. 
The latter can be computed exactly~\cite{bickelPRE04}, and we simply sketch the results
\begin{equation}
\Phi _{1}\left( z\right) =\frac{1}{\sqrt{2\pi }\xi _{\perp }}
\exp \left[ -\frac{z^{2}}{2\xi _{\perp }^2}\right] \ , \label{phi1}  
\end{equation}
for $N=1$, and
\begin{equation}
\begin{split}
\Phi _{2}&\left( {\bm\rho },{\bm\rho }';z,z'\right) =\left(
2\pi \right) ^{-1}\left[ \det \mathcal{G}\right] ^{-1/2} \\
&\times \exp \left[ -
\frac{G\left( 0\right) }{2\det \mathcal{G}}\left( z^{2}+z^{\prime 2}\right) +
\frac{G({\bm\rho }-{\bm\rho }^{\prime })}{\det \mathcal{G}}zz^{\prime
}\right] \ ,
\label{phi2}
\end{split}
\end{equation}
for $N=2$, where we define $\det \mathcal{G}=G^{2}\left( 0\right) -G^{2}({\bm\rho }-{\bm\rho }
^{\prime })$. 
Note that eq.~(\ref{u2})--(\ref{phi2})
provide an explicit criterion for the validity of the cumulant approximation.
Indeed, the dimensionless coupling constant
$\omega /\xi _{\perp }$ has to be identified as the ``small'' parameter of the cumulant
expansion. The relevance of our approach is then ensured as long as $\omega /\xi _{\perp }<1$.

\begin{figure}
\includegraphics[width=3.25in]{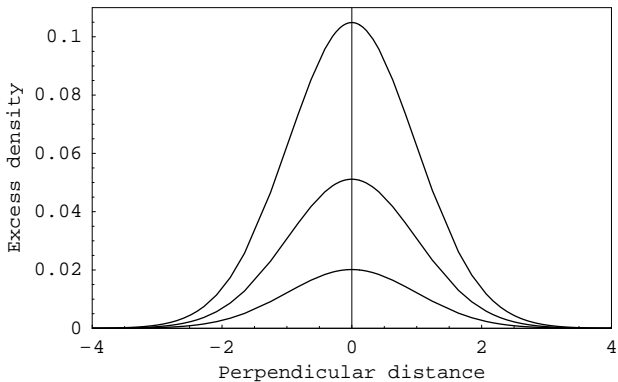}
\caption{\label{fig1} Excess particle density $\left(\rho /\rho_{\infty}-1\right)$ 
as a function of the (dimensionless) elevation $z/\xi_{\perp}$,
for different values of the surface coupling constants $\omega /\xi_{\perp}=0.1$, $0.25$
and $0.5$, respectively.}
\end{figure}

Let us now discuss some plausible outcomes of our analysis.
The density profile can be estimated from the approximate relation
$\rho \left( z\right) =  \rho _\infty \exp \left[-\beta \mathcal{U}_1\right]$,
with $\rho _\infty =  f/\lambda ^3 $ the bulk value~\cite{comment}.
Results from the last section yield
\begin{equation}
\rho \left( z\right) =\rho _\infty 
\exp \left[ \frac{\omega}{\sqrt{8 \pi}\xi _{\perp }} \exp\left[-\frac{z^2}{2\xi _{\perp }^2}\right] \right]  \ .
\label{profile}
\end{equation}
The variation of concentration upon distance is represented in Fig.~(\ref{fig1}).
Remarkably, the density reaches its bulk value over a distance 
that \emph{is not} the extrapolation length of the surface potential.
Instead, this range is set by the \emph{elastic properties} of the fluctuating 
interface via the roughness $\xi_{\perp}$.
The potential $\mathcal{U}_{1}$\ 
can be interpreted as a  ``condensation'' potential forcing the
colloids to be localized in the vicinity of the surface $z=0$.
The excess of particles is further characterized
by the \emph{adsorbance}
\begin{equation}
\Gamma = \int_{-\infty}^{+\infty} \left[\rho(z)-\rho_{\infty}\right] d z
= \frac{1}{2}\omega  \rho_{\infty}  +\mathcal{O}\left(
\omega ^2\right)  \ .
\end{equation}
As expected, the adsorbance $\Gamma$ is positive --- denoting particle accumulation --- and scales as the
coupling constant $\omega$. 
Note that in the case of colloid-surface repulsion ($\omega <0$), colloids are depleted
from the interface and the corresponding adsorbance is negative 
(see also ref.~\cite{bickelPRE04}).
Finally, let us comment on the limit of an infinitely rigid interface, corresponding to $\kappa \gg k_BT $.
From the representation of the $\delta$--distribution $\delta(x)=\lim_{\varepsilon \rightarrow 0}
(2\pi \varepsilon^2)^{-1}\exp[-x^2/(2\varepsilon^2)]$, we immediately recover the
anticipated Boltzmann distribution $\rho(z)=\rho_{\infty}\exp\left[-\beta \mathcal{H}_{cm}[0](z)\right]$, 
with the surface potential given in eq.~(\ref{intcm}).

We now focus on the two--body potential. From the general property of random variable
theory according to which $\left\langle XY\right\rangle -\left\langle
X\right\rangle \left\langle Y\right\rangle \geq 0$, one can directly
infer that $\Phi _{2}\left( {\bm\rho },{\bm\rho }^{\prime };z,z^{\prime
}\right) \geq \Phi _{1}\left( z\right) \Phi _{1}\left( z^{\prime }\right) $.
Comparing with Eq.~(\ref{u2}) allows us to 
conclude that $\mathcal{U}_{2}$ is \emph{always attractive}, whatever
the sign of $\omega$. 
Indeed, surface entropy is found to increase in every instance when particles are getting closer,
leading to fluctuation-induced attraction.

The two-body potential is shown in Fig.~(\ref{fig2}) for a fluid membrane ($\sigma=0$).
For fixed perpendicular distances $z$ and $z^{\prime }$, 
 $\mathcal{U}_{2}$ vanishes as the
relative parallel distance $l=\left| {\bm\rho }-{\bm\rho }^{\prime }\right|$
goes to infinity. When the latter is fixed, the interaction
dies off very rapidely as $z$ or $z^{\prime }\gg \xi_{\perp}$.
From Eq.~(\ref{u2})--(\ref{phi2}), it can be noticed that $\mathcal{U}_{2}$ is actually boundless
when $l \rightarrow 0$. For two particles at the same elevation $z=z'=0$ we find
\begin{equation}
\beta \mathcal{U}_{2}\left(  l \right)  \sim -\frac{1}{16 \pi}
\left( \frac{\omega }{\xi_{\perp}} \right)^2 \left( 1 -\frac{G\left(l\right)^2}
{G\left(0\right)^2} \right)^{-1/2}  \ ,
\label{shortu2}
\end{equation}
where the Green function $G(l)$ depends on the elastic properties of the surface
under consideration. As a matter of fact, this behaviour is reminiscent 
of the short-range colloid-surface potential.
In a real system, however, particles always have a finite size so 
that a minimal distance  between colloids is set by their diameter. 
Moreover, because $G(l)$ usually involves logarithmic contributions at short distances,
the attraction is expected to increase only very slowly.  It thus remains moderate even at low separations. 

So far, we have not specified the nature of the surface and of the inter-particle potential 
$\mathcal{H}_{cc}$.
To keep the discussion simple, we describe the colloids as hard spheres of diameter $d$. 
We consider as a first example an interface between two immiscible liquids,
with no bending rigidity. 
The height correlation function is
$G(l)=(2\pi \beta \sigma)^{-1} \mbox{K}_0 ( l /\xi ) $,
with $\xi = (\sigma/\mu)^{1/2}$ the capillary length. At short distances, 
the Bessel function $\mbox{K}_0$ behaves like $\mbox{K}_0(x)\sim_0 -\ln (x)$ and
the mean square fluctuations of the interface involves some molecular size $\lambda_c$~\cite{safranbook}. 
With $\lambda_c/\xi=10^{-4}$,  $d/\xi=10^{-3}$,
and $\omega /\xi_{\perp}=0.5$, we obtain the value at contact $ \mathcal{U}_{2}(d)\approx 5
\times 10^{-3}k_BT$. 

In the opposite limit of a fluid membrane with vanishing surface tension $\sigma =0$,
the height fluctuations are much larger and surface-mediated attraction is expected to be stronger. 
The short-distances behaviour of the propagator that appears in eq.~(\ref{phi2}) is
\begin{equation}
G\left( l \right) \simeq G\left( 0\right) \left( 1+\frac{2}{\pi }\frac{
l ^{2}}{\xi _{\parallel }^{2}}\ln \left( \frac{l }{\xi _{\parallel }}
\right) \right)  \quad \text{for} \quad l \ll \xi _{\parallel } \ ,
\label{greenmembrane}
\end{equation}
with $\xi _{\parallel }=( 4\kappa /\mu ) ^{1/4}$ the in-plane correlation length of the membrane.  
From eq.~(\ref{shortu2}) we obtain the potential ``at contact''
\begin{equation}
\beta \mathcal{U}_{2} \left( | {\bm r }-{\bm r }^{\prime }|=d \right) \approx - \frac{1}{32\sqrt{\pi}}
\left(\frac{\omega }{\xi_{\perp}}\right)^2 \left[  \frac{d^2}{\xi_{\parallel}^2}\ln
\left(\frac{\xi_{\parallel}}{d} \right)          \right]^{-1/2}  \ .
\label{shortu2mem}
\end{equation}  
For sensible values of the parameters $d/\xi_{\parallel}=10^{-3}$ and $\omega /\xi_{\perp}=0.5$, 
the depth of the potential well is comparable to the thermal energy: 
$ \mathcal{U}_{2}(d) \approx 1.7 k_BT$. 
Actually, the strength of the interaction increases as the ratio $d/\xi_{\parallel}$ decreases.
The only requirement limiting the lower bound of this ratio 
is that we have to remain in the \emph{colloidal domain},
\emph{i.e.} the diameter of the colloids has to be large compared to the size of the solvent molecules.

\begin{figure}
\centering
\includegraphics[width=3.25in]{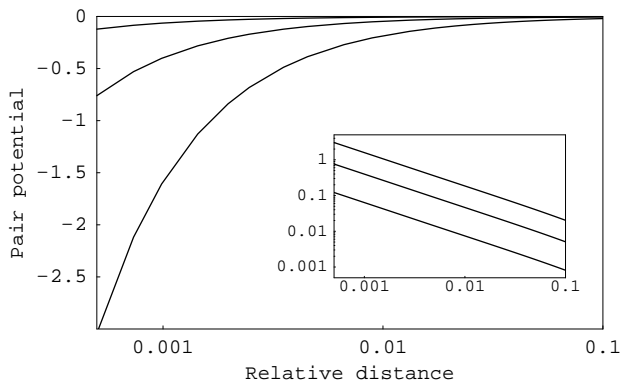}
\caption{\label{fig2}
Effective two-body potential (in units of $k_BT$) as a function of the relative 
distance $l /\xi_{\parallel}$, for a fluid membrane ($\sigma=0$), 
and for two colloids at the same elevation $z=z'=0$. Note the logarithmic scale on the horizontal axis. 
The different curves correspond to  surface coupling constants $\omega /\xi_{\perp}=0.1$, $0.25$
and $0.5$, respectively. In the inset, we show $\ln |\beta \mathcal{U}_{2} |$ as a function of
$\ln ( l / \xi_{\parallel} ) $, illustrating eq.~(\ref{shortu2mem}).}
\end{figure}

The original motivation of our work was to describe the statistical properties 
of colloidal suspension in contact with a soft surface.
The main outcome of this article is that, as a result of surface fluctuations, 
particles experience a potential of mean force $\mathcal{U}
=\mathcal{U}_1+\mathcal{U}_2+ \ldots$, that is not pairwise additive. Indeed,  the expectation value 
eq.~(\ref{defmeanforce}) involves three--body contributions and higher,
but those are neglectable at low density or for small surface coupling parameter.
The leading contribution, $\mathcal{U}_1$, act as an external field
that condensates the colloids in the vicinity of the interface.
In the weak adsorption regime, particle accumulation remains moderate:
for $\omega/\xi_{\perp}=0.5$, the density at elevation $z=0$ 
is only 10~$\%$ higher than its bulk value.
But once localization is enforced, particles attract each other 
through the pair potential $\mathcal{U}_2$
whose strength can be quite substantial. Under some circumstances, 
it is of the order of the thermal energy $k_BT$.
Accordingly, these forces may have a strong influence on the local organization of the suspension,
and surface fluctuations might even possibly
promote the nucleation of an ordered phase of colloids in the vicinity
of the interface.
This suggestion would clearly deserve more attention
as it may provide a new route for directed self-assembly of mesoscopic
colloidal structures.

Finally, we note that the mechanism leading to particle attraction 
does not seem to depend crucially on the exact nature
of $\mathcal{H}_{cm}$.
The effective potential is solely driven by the entropy of the surface
and shares many similarities with the well-known depletion forces.
In particular, the dependency of $\mathcal{U}_2$ on the elastic parameters 
$\kappa$, $\sigma$ and $\mu$ turns out to be a great advantage as
it may allow to \emph{tune} the interaction. This can be achieved for instance
by adding cosurfactant molecules to monolayers and bilayers, 
or by slightly changing the temperature
for an interface between two immiscible liquids \emph{near the mixing transition}. Indeed,  
the surface tension scales in the latter situation as $\sigma \sim |T-T_c|^{-\mu}$, with $\mu >0$ a critical exponant,
in such way that the fluctuations amplitude is extremely sensitive to a minute variation of temperature.
It should therefore be possible, in principle, to control the effective interactions
between colloids.

\end{document}